\documentclass[aip,apl,a4paper, reprint,amsmath,amssymb,floatfix]{revtex4-1}

\usepackage{bm}
\usepackage{amssymb}
\usepackage{graphicx}
\usepackage{SIunits}
\usepackage{soul}

\usepackage[normalem]{ulem}

\renewcommand{\exp}[1]{\ensuremath{\mathrm{e}^{#1}}}
\renewcommand{\vec}[1]{\ensuremath{\bm{#1}}}
\newcommand{\abs}[1]{\ensuremath{|#1|}}
\newcommand{\mum}[1]{\unit{#1}{\micro\meter}}
\newcommand{\nm}[1]{\unit{#1}{\nano\meter}}
\newcommand{\rcm}[1]{\unit{#1}{\reciprocal{\centi\meter}}}
\newcommand{\ps}[1]{\unit{#1}{\pico\second}}

\newcommand\trick[1]{}

\begin{document}

\title{Differential ultrafast all-optical switching of the resonances of a micropillar cavity}

\author{Henri Thyrrestrup}\email{h.t.nielsen@utwente.nl, website: www.photonicbandgaps.com}
\author{Emre Y\"uce}
\author{Georgios Ctistis}
\affiliation{Complex Photonic Systems (COPS), MESA+ Institute
for Nanotechnology, University of Twente, 7500 AE Enschede, The Netherlands}

\author{Julien Claudon}
\affiliation{Univ. Grenoble Alpes, INAC-SP2M, Nanophysics and Semiconductors Lab, F-38000 Grenoble, France}
\affiliation{CEA, INAC-SP2M, Nanophysics and Semiconductors Lab, F-38000 Grenoble, France}

\author{Willem L. Vos}
\affiliation{Complex Photonic Systems (COPS), MESA+ Institute
for Nanotechnology, University of Twente, 7500 AE Enschede, The Netherlands}

\author{Jean-Michel G\'erard}\email{jean-michel.gerard@cea.fr}
\affiliation{Univ. Grenoble Alpes, INAC-SP2M, Nanophysics and Semiconductors Lab, F-38000 Grenoble, France}
\affiliation{CEA, INAC-SP2M, Nanophysics and Semiconductors Lab, F-38000 Grenoble, France}

\date{\today}


\begin{abstract}
We perform frequency- and time-resolved all-optical switching of a GaAs-AlAs micropillar cavity using an ultrafast pump-probe setup. The switching is achieved by two-photon excitation of free carriers. We track the cavity resonances in time with a high frequency resolution. The pillar modes exhibit simultaneous frequency shifts, albeit with markedly different maximum switching amplitudes and relaxation dynamics. These differences stem from the non-uniformity of the free carrier density in the micropillar, and are well understood by taking into account the spatial distribution of injected free carriers, their spatial diffusion and surface recombination at micropillar sidewalls.
\end{abstract}

\maketitle

Micropillar cavities are versatile solid-state nanophotonic structures that locally enhance the light-matter interaction due to their high quality-factors and small mode volumes.\cite{gerard1998aa,bayer2001aa,Solomon2001} Moreover, they provide a clean free-space optical interface with nearly perfect in- and out-coupling efficiency. These qualities have stimulated the successful application of micropillar cavities with embedded quantum dot emitters as efficient single photon sources\cite{Moreau2001,Pelton2002,Gazzano2013} and diode lasers\cite{Jewell1991}, and for observing cavity QED strong coupling.~\cite{Reithmaier2004,Loo2010} In all these realizations, however, the cavity resonance is stationary in time, certainly during relevant interaction times such as an emitter lifetime. To give micropillar cavities new functionality, we propose to bring micropillars into a new dynamic regime where the cavity resonance is switched on a timescale faster than the emitter lifetime.\cite{Thyrrestrup2013,Fiore2013} To this aim we have studied the ultrafast dynamics of all-optically switched micropillar cavities. 

Micropillar cavities support multiple transverse localized modes with distinct mode profiles. We identify multiple transverse resonances of simple micropillar cavities and perform frequency and time-resolved switching of the resonances via all-optical excitation of free carriers. Compared to previous switching experiments on micropillars\cite{Jewell1989}, we here study the temporal dynamics of several resonances. The different transverse modes in the micropillar cavity shift in frequency by different magnitudes and show different switching dynamics, pointing to a very significant role of the inhomogeneous spatial distribution of the free carriers in the micropillar. 

\begin{figure}[tb]
\centering
\includegraphics[width=8cm]{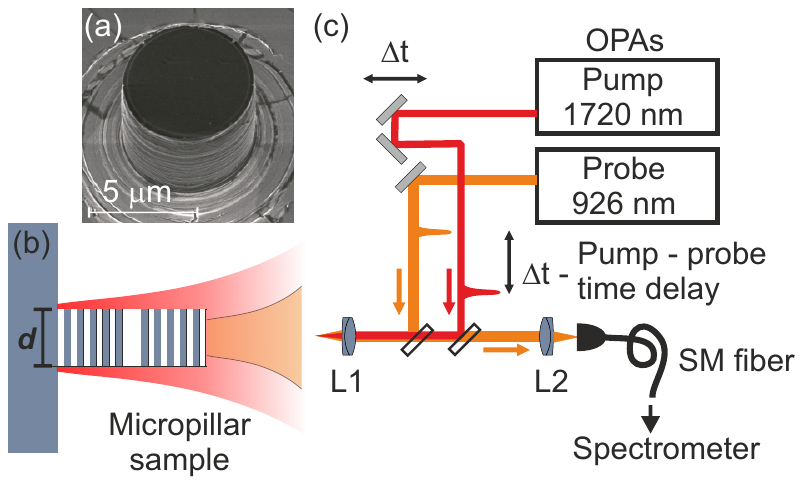}
\caption{(color online) (a) SEM image of a micropillar with a diameter of $d=\mum{6}$, composed of two GaAs-AlAs Bragg mirrors separated by a GaAs $\lambda$-layer. (b) Schematic of the cavity with the probe beam (orange) and the a larger pump (red) beam. (c) Ultra-fast pump-probe setup. The pump and probe pulses are generated by two synchronized  optical parametric amplifiers (OPAs) with a repetition rate of \unit{1}{\kilo\hertz}. A delay stage controls the time delay $\Delta t$ between the two pulses, which are focused on the top facet of the micropillar by the achromatic lens L1. Reflected probe light is collected by a single mode (SM) fiber through L2 and sent to a spectrometer with an InGaAs diode array.
}
\label{fig:setup} 
\end{figure}

The micropillar cavities as shown in Fig.~\ref{fig:setup}(a) are etched from a layered structure that consists of GaAs-AlAs Bragg stacks separated by a central GaAs $\lambda$-layer, grown using molecular-beam epitaxy. The bottom Bragg stack consists of 25 pairs of GaAs-AlAs layers and the top contains 13 pairs. The GaAs and AlAs layers are \nm{65.4} and \nm{78.3} thick, respectively, and the central GaAs $\lambda$-layer has a thickness of \nm{261.7}. The resulting design wavelength of the planar cavity structure is \nm{916}. The planar structure has been etched by reactive ion-etching resulting in free standing micropillars with a range of diameters between \mum{1} and \mum{20}. During etching, a thin layer of $\mathrm{SiO}_x$ with a thickness between 100 and 200 nm is formed on the sidewalls to prevent oxidation of the AlAs layers. Switching experiments have been performed on pillars with a diameter of \mum{6}, which have a suitable size to reveal the effects of a spatially dependent free carrier density. \footnote{We have observed qualitatively similar results on pillars with $d=\mum{3,20}$, although the temporal range (\ps{100}) was too limited for modelling.}

\begin{figure}[tb]
\centering
\includegraphics[]{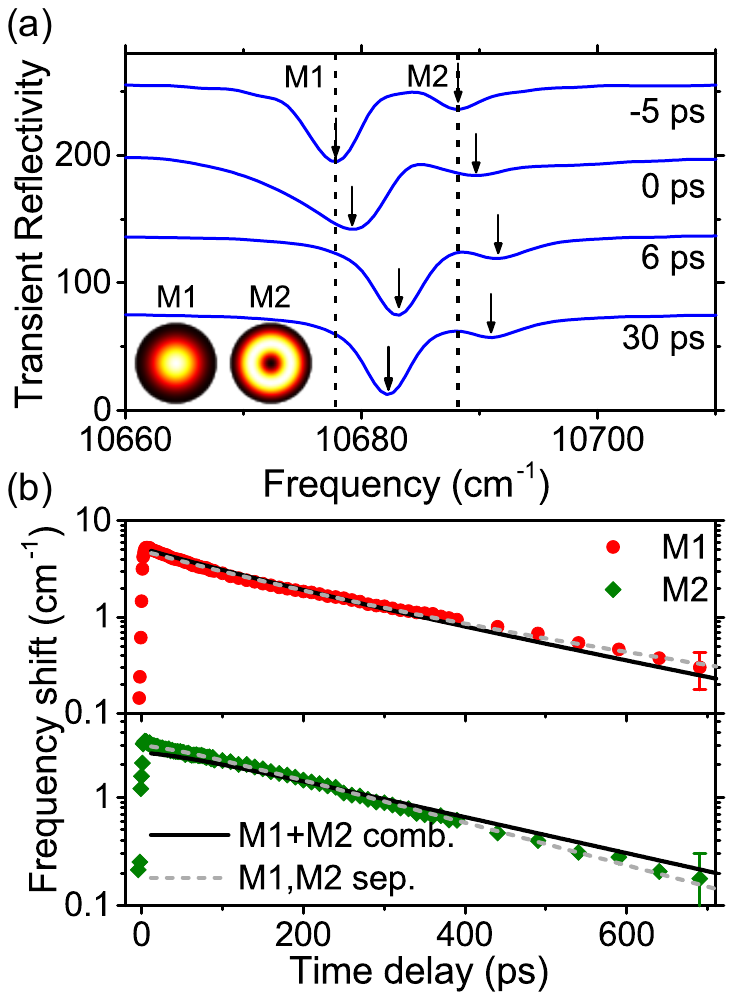}
  \caption{\label{fig:measurements} (a) Transient reflectivity spectra at different pump-probe time delays. The two lowest order resonances M1 and M2 are seen as troughs. Both resonances shift to higher frequencies  by up to \rcm{5.2} and \rcm{3.3}, respectively, at $\Delta t=\ps{6}$. (b) Frequency shift of the two resonances M1 and M2 versus pump-probe  delay. The lines model the decay using a diffusion model for the free carriers. Dashed lines are seperate fits with $\tau_\mathrm{M1
}=\ps{369\pm 40}$, $S_\mathrm{M1}\approx\unit{0}{\centi\meter\per\second}$ and $\tau_\mathrm{M2}=\ps{280\pm 55}$, $S_\mathrm{M2}=\unit{3\times 10^5}{\centi\meter\per\second}$.}
\end{figure}

The transverse cavity modes of a micropillar cavity can be classified in terms of the mode indicies for the propagating modes in an infinite circular waveguide \cite{Marcuse1991} and a quantized propagation constant. The propagation constant is determined by the longitudinal resonance condition with an effective refractive index \cite{Gerard1996a}, where each mode has an associated spatial mode profile. The first two modes are degenerate $\mathrm{HE}_{11}$ waveguide modes with orthonormal polarizations, both with a Gaussian profile, see Fig.~\ref{fig:measurements}(a). These two form a single resonance that we denote M1. The following four mode corresponds to the $\mathrm{TE}_{01}$, the two degenerate $\mathrm{HE}_{21}$, and the $\mathrm{TM}_{01}$ modes. These four modes are nearly degenerate with a ring-shaped intensity profile as shown in Fig.~\ref{fig:measurements}(a). Due to this similarity and the small spectral splitting relative to their linewidth we consider them as a single resonance that we denote M2. \cite{Ctistis2010} Higher order modes are not detected in the reflectivity experiment due to spatial filtering and thus are not considered further.

The micropillar cavities are switched with the ultrafast pump-probe setup shown in Fig.~\ref{fig:setup}. A \unit{150}{\femto\second} short pump pulse with an pump pulse energy of $E=\unit{1.2}{\nano\joule}$ excites free carriers in the GaAs layers. The free carriers decrease the refractive index \cite{Mondia2005} and thereby blue shift the resonance frequency of the cavity modes. \cite{Fushman2007b,Harding2007,Tanabe2009b,Nozaki2010}
 The pump frequency $\tilde\nu_\mathrm{pu}=\rcm{5814}$ ($\lambda_\mathrm{pu}=\nm{1720}$) is tuned above half the band gap energy of GaAs to ensure a homogeneous free carrier excitation along the micropillar via two-photon absorption. \cite{Euser2005} The frequency shift of each cavity resonance is detected by a \unit{150}{\femto\second} probe pulse whose timing is set by a delay stage. The probe is focused on the micropillar and is centered by maximizing the total reflected intensity. The pump spot has a Gaussian radius of $w_0 = \mum{3.5}$ and is centered on the pillar by maximizing the switching magnitude of the M1 resonance. We find that small displacements in the pump position from the center drastically alter both the switching magnitude and the visibility of higher order resonances. The reflected probe light is spatially filtered by a single-mode fiber and is dispersed in a spectrometer with a resolution of $\delta\tilde\nu\approx \rcm{1}$. The center frequency of the probe is $\tilde\nu_\mathrm{pr}=\rcm{10675}$ ($\lambda_\mathrm{pr}=\nm{936}$) with a bandwidth of \rcm{175}\! covering the full spectrum of all the cavity resonances. The transient reflectivity is obtained by normalizing the time integrated probe intensity to the reflectivity spectrum of a gold mirror. The time integrated intensity includes light that has been stored in the cavity for approximately \unit{500}{\femto\second}. \cite{Euser2009}

The cavity resonances are tracked in time by recording the transient reflectivity spectra versus pump-probe delay as shown in Fig.~\ref{fig:measurements}(a). The unswitched reflectivity spectrum at negative time delays $\Delta t < 0$ shows two resonances M1 and M2 at $\nu_{M1} = \rcm{10677.8}$ and $\nu_{M2} = \rcm{10688.2}$ that agrees well with previous linear reflectivity measurements \cite{Ctistis2010}. The difference in visibility of the two troughs is a result of the different overlap between the intensity profiles of the individual cavity modes and the probe pulse. The deep M1 trough with a visibility of 0.8 shows the good mode match between the probe beam, and the Gaussian-like mode for this resonance. The spatial overlap between the Gaussian probe and the ring-shaped modes is much smaller, evident in the shallow M2 trough. 

At zero time delay ($\Delta t=\ps{0}$) the pump pulse excites electron-hole pairs in the GaAs layers, which decreases the refractive index linearly with the free carrier density, shifting both resonances to higher frequencies. The broadening of the two troughs at $\Delta t=0$, when pump and probe exist simultaneously, is attributed to non-degenerate free carrier absorption of one pump and one probe photon.\cite{Harding2007} The maximum frequency shift of $\Delta\tilde\nu_\mathrm{M1}=\rcm{5.2}$ and $\Delta\tilde\nu_\mathrm{M2}=\rcm{3.3}$ is seen at $\Delta t\approx\ps{6}$, after the free carriers have thermalized. \cite{Huang1998,Harding2007} Following the excitation, the free carriers recombine and the resonances return to their original frequencies. The most important observation in Fig.~\ref{fig:measurements}(a) is the difference in the frequency shifts of the M1 and M2 resonances at $\Delta t\approx\ps{6}$. A homogeneous carrier distribution would result in an equal shift of the two resonances. The difference in maximum frequency shift is therefore attributed to an initial spatial dependent free carrier density that has a stronger overlap with the modes associated with resonance M1. 

The frequency shift of the two resonances is studied in the range  between $\Delta t=\ps{-5}$ and \ps{700}. The high resolution time traces shown in Fig.~\ref{fig:measurements}(b) allow us to obtain detailed information on the relaxation dynamics of the free carriers. The resonance frequencies are extracted with subpixel resolution and we estimate the uncertainty in the frequency shifts to be better than $\pm\rcm{0.125}$. Both curves deviate from single-exponential decay. The relaxation dynamics for M1 is convex on a log scale with an initial faster decay. The M2 resonance on the other hand has an overall concave shape with initially a slower decay that gradually speeds up. Thus the relaxation dynamics of the two resonances differs qualitatively within $\Delta t\lesssim \ps{200}$. At longer times the decay is exponential for both curves, with a slightly faster decay rate for M2.

\begin{figure}[tb]
 \includegraphics[]{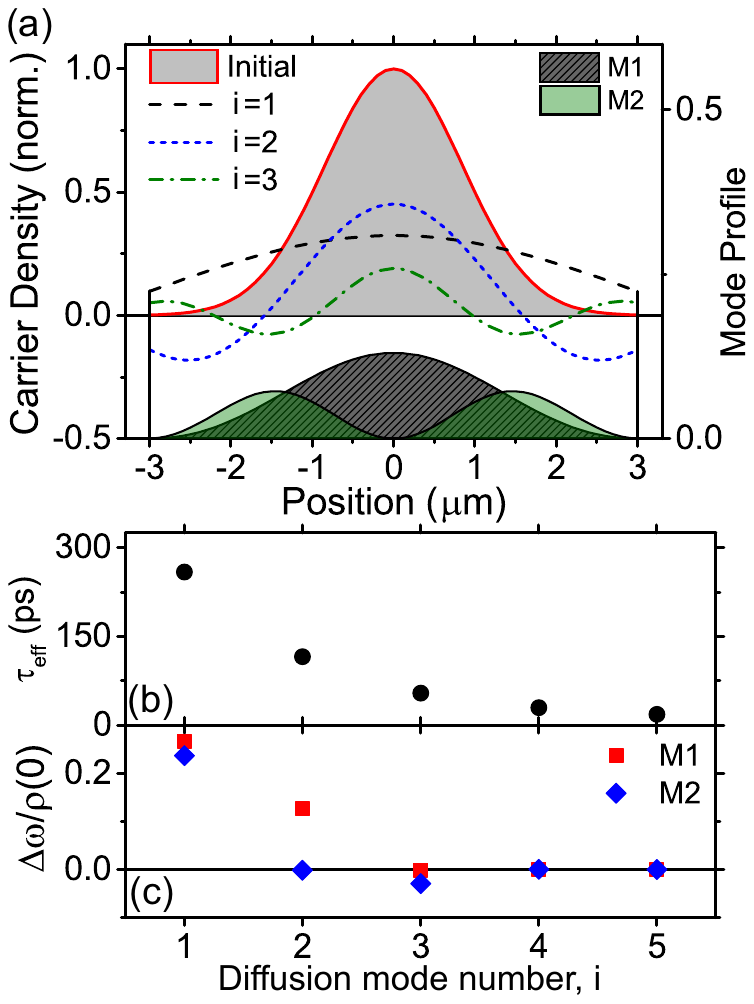}
  \caption{\label{fig:CarrierDist} (a) The initial free carrier distribution $n_0(r)$ decomposed into modes of the solution to the diffusion equation for $\tau=\ps{345}$ and $S=\unit{3\times 10^5}{\centi\meter\per\second}$. Below are the normalized electromagnetic profiles of the M1 and M2 resonances. All curves are cross sections at the center of the $\lambda$-layer. The spatial shape of the initial carrier density corresponds to the square of the HE11 waveguide mode intensity for $\lambda=\nm{1720}$. (b) The effective decay time $\tau_\mathrm{eff}$ for the diffusion modes. (c) The initial contribution to the switching magnitude of M1 and M2 for the different diffusion modes.}
\end{figure}

For a low free carrier density the relaxation is governed by exponential decay describing the nonradiative trapping of single carriers. However, a single linear relaxation rate of the carriers alone cannot describe the different dynamics of the resonances. To account for this difference we consider the spatial free carrier dynamics, since the frequency shift $\Delta\omega_i(t)$ of individual resonances is given by the overlap \cite{Joannopoulos2008} between the carrier distribution $n(\vec r)$ and the mode profile $\abs{\vec E
_i(\vec r,t)}^2$, 
\begin{equation}
 \frac{\Delta\omega_i(t)}\omega \propto \int \varepsilon(\vec r)\abs{\vec E
_i(\vec r,t)}^2n(\vec r,t)\,d^3\vec r.\label{eq:freqshift}
\end{equation}
The spatial dynamics is governed by ambipolar diffusion of electron-hole pairs in the Bragg layers and the relaxation at recombination sites in the pillar and at the surface. Since the initial carrier density is homogeneous throughout the length of the pillar \footnote{We estimate that the pump intensity is depleted by less than 10\% by propagating through the pillar structure.} we only consider the radial dependence. Over time the carriers will recombine at the GaAs-AlAs interface and the the validity of our assumption relies on the thin GaAs sublayers, and that the interface recombination velocity is slow.  The solution to the time-dependent ambipolar diffusion equation for the carrier density $n(\vec r,t)$ with absorbing boundary conditions $\vec e_r\cdot D\,\nabla n(\vec r) = - S n(\vec r)$ can be written analytically in cylindrical coordinates as:
\begin{equation}
n(r,t)=\frac 2{R^2}\sum_{i=1}^\infty \exp{-(\alpha_i^2 D+1/\tau) t}\frac{\alpha_i^2 J_0(\alpha_i r)}{(h^2+\alpha_i^2)J_0^2(\alpha_i R)} A_i,\label{eq:diffsol}
\end{equation}
where $R=d/2$ is the pillar radius, $\alpha_i$ the positive roots of the transcendental Bessel equation $\alpha J_1(\alpha R)-h J_0(\alpha R)=0$, $h=S/D$, and $A_i=\int_0^R n_0(r) J_0(r\alpha_i)r\, dr$ is a coefficient that depends on the initial free carrier density $n_0(r)$. \cite{Carslow1959} The pillar acts as a waveguide for the pump pulse. For our pumping geometry, the pump pulse is coupled predominantly to the fundamental mode, so that we assume the initial carrier density shown in figure 3 to be proportional to the squared intensity of the HE11 waveguide mode at $\lambda = \nm{1720}$. The diffusion constant is $D=\unit{25}{\centi\meter^2/s}$ \cite{Ruzicka2010} and $S$ the surface recombination velocity and $\tau$ is a phenomenological bulk recombination rate that also includes the recombination at the GaAs/AlAs interfaces. 

The dashed lines in Fig.~\ref{fig:measurements}(b) are fits to Eq.~\eqref{eq:freqshift} and Eq.~\eqref{eq:diffsol} with $S$, $\tau$ and an amplitude as free parameters. The optimization is performed by  minimizing the difference area squared between model and data to include different sampling intervals. Remarkably, both fits replicate the data well, including the initial convex and concave dynamics of the two curves, which confirms that the initial dynamics is indeed governed by spatial diffusion of the free carriers. The fits constrain $S$ poorly with an uncertainty of several orders of magnitude, and is mainly included for reference. If we force common material parameters for the two curves the solid lines represent combined fits with a single amplitude and gives $\tau=\ps{345\pm 45}$, $S=\unit{2.8\times 10^5}{\centi\meter\per\second}$. The difference in initial frequency shift is understood since the initial carrier density $n_0(r)$ overlaps well with the M1 profile (Fig.~\ref{fig:CarrierDist}(a)) and much less with the M2 profile. The amplitude corresponds to an initial maximum free carrier density of $n_0(0)=\unit{2.1\times 10^{-18}}{\centi\meter\rpcubed}$.

To interpret the dynamics we note that in Eq.~\eqref{eq:diffsol} the free carrier density $n(r,t)$ can be interpreted as a sum of terms with different spatial profiles. Each term ($i$) decays in time with a different decay time $\tau_\mathrm{eff}=1/(\alpha_i^2 D+1/\tau)$ shown in Fig.~\ref{fig:CarrierDist}(b). The decay time is a nonanalytical combination of the three parameters, $D$, $\tau$ and $S$. For the selected parameters the first three terms decay within $\tau_1=\ps{260}$, $\tau_2=\ps{115}$ and $\tau_3=\ps{55}$, respectively. Their contribution  to the frequency shifts of the electromagnetic resonances is shown in Fig.~\ref{fig:CarrierDist}(c). The interesting spatial dynamics that separates the different resonances happens within the first \ps{300} where the first three terms contribute significantly to the frequency shift. The HE11 mode has a positive overlap with the second term and almost zero overlap with the third, which leads to the initial faster decay of the M1 resonance. For the M2 resonance the contribution from the second term is zero since the antinodes of the HE21 mode coincide with the nodes of the free carrier density contribution. This leaves only a negative short lived contribution from the third term resulting in the slower decay in the beginning. In simple words, at short times the carriers quickly diffuse from the M1 profile and overlap more with the M2 profile, resulting in different shapes for the decay curves. At long times, only the slowest first term survives, resulting in a single exponential decay and the shape of the free carrier density converges to the nearly flat shape for $i=1$. This term has a near identical overlap with both the HE11 and HE21 modes. 

An alternative mechanism to explain the nonexponential decay is to consider higher order decay terms where the second order is the  bimolecular electron-hole radiative recombination. \cite{tHooft1981,Reinhart2006} However, fitting the M1 curve in Fig~\ref{fig:measurements}(b) to such a model gives a radiative recombination coefficient of $B=\unit{6.9\times 10^{-9}}{\centi\meter\cubed\per\second}$, which is one to two orders of magnitude larger than the established value for GaAs of $B\approx\unit{1\times 10^{-10}}{\centi\meter\cubed\per\second}$. In addition, it does not explain the qualitatively different relaxation curves, and especially not the concave shape for M2. We therefore believe that a position-dependent carrier density is the most likely explanation.

In summary, we have presented ultrafast all-optical switching of GaAs/AlAs micropillar cavities. As the switching magnitude of each cavity resonance is determined by the spatial overlap between the free carrier density and the mode profile, different modes exhibit different frequency shifts and relaxation dynamics. By adaptively exciting certain carrier-density volumes, it will in the future become possible to achieve dynamic and selective control of distinct cavity resonances in time.

This work was partly funded by the FOM "Zap!" project, which is financially supported by NWO and by NWO-Nano, and by STW. We thank an anonymous referee for excellent suggestions.

%


 \end{document}